\documentclass[twocolumn,showpacs,preprintnumbers,amsmath,,amssymb,aps,prc]{revtex4}
\usepackage{graphicx}% Include figure files
\usepackage{dcolumn}% Align table columns on decimal point
\usepackage{bm}% bold math

\begin{document}

\preprint{APS/123-QED}

\title{\Large Systematic study of projectile structure effect  on fusion barrier distribution\\}% Force line breaks with \\

\author{Pratap Roy$^{1}$}
\email{pratap_presi@yahoo.co.in} 
\author{A. Saxena$^1$} 
\author{B.K. Nayak$^1$}
\author{E.T. Mirgule$^1$} 
\author{B. John$^1$}
\author{Y.K. Gupta$^1$} 
\author{L.S. Danu$^1$} 
\author{R.P. Vind$^1$} 
\author{Ashok Kumar$^2$} 
\author{R.K. Choudhury$^1$}
%\author{}%
 
\affiliation{%
$^1$Nuclear Physics Division, Bhabha Atomic Research Centre, Mumbai - 400085, INDIA
 \\
$^2$School of Basic and Applied Sciences, Shobhit University, Meerut. U.P. India.250009}%

\author{}
% \homepage{}
\affiliation{
\\
 forced% with \\
}%

\date{\today}% It is always \today, today,

\begin{abstract}
Quasielastic excitation function measurement has been carried out for the $^{4}$He + $^{232}$Th system at $\theta_{lab}$=160$^\circ$ with respect to the beam direction, to obtain a representation of the fusion barrier distribution.  Using the present data along with previously measured barrier distribution results on  $^{12}$C, $^{16}$O, and $^{19}$F + $^{232}$Th systems a systematic analysis has been carried out to investigate the role of target and/or projectile structures on  fusion barrier distribution.
It is observed that for $^{4}$He, $^{12}$C, and $^{16}$O + $^{232}$Th, reactions the couplings due to target states only are required in coupled channel fusion calculations to explain the experimental data, whereas for the $^{19}$F+ $^{232}$Th system along with the coupling of target states, inelastic states of  $^{19}$F are also required to explain the experimental results on fusion-barrier distribution. The width of the barrier distribution shows interesting transition behavior when plotted with respect to the target-projectile charge product for the above systems.

% \verb+\pacs{25.70.Mn, 25.70.Bc, 24.10.Eq}+ .
\end{abstract}

\pacs{25.55.Ci, 25.70.Bc, 25.70.Jj}% PACS, the Physics and Astronomy
\maketitle

%\section{\label{sec:level1}Introduction:\protect\\}

Heavy-ion fusion reaction at low energies is generally described as a one-dimensional barrier penetration problem, in which the radial motion is the only degree of freedom involved in the fusion process. The model based on this framework is known as the one-dimensional or single-barrier penetration model. For incident energies well above the Coulomb barrier, the measured fusion cross sections are well reproduced by the one-dimensional barrier penetration model. However at near and below Coulomb barrier energies, it has been observed that the experimental fusion cross section for many systems is much higher than the prediction of this model~\cite{1,2,3}. This phenomenon of  enhancement of sub-barrier fusion cross section has been interpreted in terms of couplings of target and/or projectile intrinsic degrees of freedom, such as static deformation, inelastic excitation, and nucleon transfer to the relative motion~\cite{4,5}.
The coupling gives rise to a distribution of fusion barriers, and passage over the lower barriers is responsible for the fusion enhancement at the sub-barrier energies. The fusion barrier is represented by a distribution (D(B)), such that the total fusion cross-section is given by,  
\begin{equation}
%\label{eq_temp}
\sigma^{fus}(E) = \int^{\infty}_0 D(B) \sigma^{fus}(E,B)\,dB , 
\end{equation} 
where the distribution D(B) is a weighting function with,
\begin{equation}
%\label{eq_temp}
\int^{\infty}_0 D(B) \,dB = 1. 
\end{equation} 

The fusion barrier distribution is defined as,
\begin{equation}
%\label{eq_temp}
D(B) =\frac{dT_0}{dE} = -\frac {dR_0}{dE}, 
\end{equation}
where $T_0$ and $R_0$ are the transmission and the reflection coefficients, respectively, for angular momentum $\ell$ = 0. It has been shown that the fusion-barrier distribution can be extracted experimentally from the fusion excitation function measurement~\cite{6} by 
\begin{equation}
%\label{eq_temp}
D_{fus}(E) = \left(\frac{1}{\pi {R_f}^2}\right) \frac{d^2}{dE^2}[E \sigma_{fus}(E)], 
\end{equation}
or from quasi-elastic-scattering measurement~\cite{7} by
\begin{equation}
%\label{eq_temp}
D_{qel}(E) = - \frac{d}{dE} \left[\frac{d\sigma_{qel}(E)}{d\sigma_R(E)}\right],
\end{equation}

where, $R_f$, $\sigma _{fus}$, $\sigma _{qel}$, $\sigma _{R}$, and E  are the barrier radius, fusion cross-section, quasi elastic scattering cross section, Rutherford scattering cross section and center of mass energy, respectively.\\

Since fusion is related to the transmission through the barrier for $\ell$ = 0, whereas large-angle quasi-elastic scattering is related
to reflection at the same barrier, these two processes
are complementary to each other. It has been shown that general features of the fusion-barrier distribution remain the same in the two representations~\cite{8,9,10}. However, from the measurement point of view quasi-elastic scattering is usually much simpler to investigate experimentally
than fusion. 
Although experimentally derived barrier distributions give valuable information on the structure of target and projectile nuclei in terms of coupling of various intrinsic degrees of freedom to relative motion, the identification of the dominating channels that act as the main doorway to the fusion is still a challenging task.
In order to identify the role of target and/or projectile structure on fusion-barrier distributions and to find the relative importance of various channel couplings, barrier distribution for the reaction $^{4}$He+$^{232}$Th has been measured and along with the previously measured  results on $^{12}$C, $^{16}$O
and $^{19}$F+$^{232}$Th systems~\cite{11,12}, a systematic analysis of the fusion barrier distributions has been carried out. In the past, projectile structure effects on quasi-elastic barrier distributions have been studied for $^{20}$Ne+$^{90,92}$Zr systems \cite{13}. For $^{20}$Ne+$^{90}$Zr system expected barrier structures due to highly deformed $^{20}$Ne projectiles have  been observed;
however, for the $^{20}$Ne+$^{92}$Zr system, smearing of barrier distribution has been reported due to scattering into noncollective inelastic channels. \\

The fusion-barrier distribution for the $^{4}$He + $^{232}$Th system should have only the target structure effect as the projectile $^{4}$He is a closed shell nucleus having the first excited state around 20 MeV. It is possible to fix target intrinsic properties by comparing coupled channel predictions with the experimental fusion-barrier distribution for the $^{4}$He + $^{232}$Th system.
Once the target intrinsic structure parameters are fixed, it is possible to investigate projectile structure effects on fusion-barrier distributions of $^{12}$C, $^{16}$O, and $^{19}$F + $^{232}$Th systems. \\

\begin{figure}[t]
\includegraphics[scale=0.63]{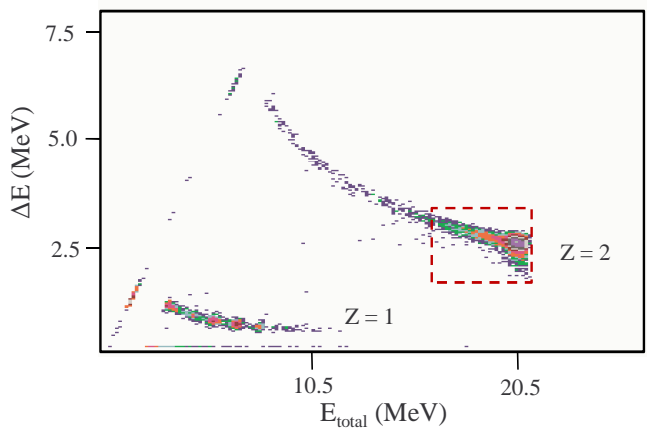}% Here is how to import jpg/ png/ pdf figure
\caption{\label{fig1} (Color online)  A typical $\Delta$E vs E scatter plot for the $^{4}$He + $^{232}$Th system.}
\end{figure}
\begin{figure}
\includegraphics[scale=0.52]{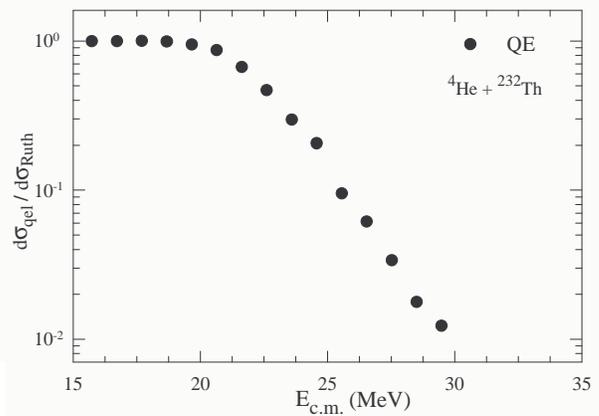}% Here is how to import jpg/ png/ pdf figure
\caption{\label{fig2}  Quasi-elastic-scattering excitation function for the $^{4}$He + $^{232}$Th system measured at $\theta_{lab}$ = 160$^\circ$. }
\end{figure}

%\section{\label{sec:level2}Experimental Details}

The experiment was performed with $^{4}$He beam from the 14 UD BARC-TIFR Pelletron accelerator facility, Mumbai, India. A self-supporting $^{232}$Th target of thickness $\sim$ 2 mg/cm$^2$ was bombarded with the alpha particles in the energy range $E_{lab}$ = 16 to 30 MeV in steps of 1.0 MeV. The energy loss by the beam particles in the half thickness of the target varies between 120 keV and 82 keV for the incident energy range of 16 to 30 MeV, which has been taken care of in the analysis. A $\Delta$E (50 $\mu$m) - E (150 $\mu$m) silicon
surface barrier detector telescope was mounted at an angle of 160$^\circ$ to the beam direction to detect the outgoing particles. Another silicon surface barrier detector at
an angle of 20$^\circ$ with respect to the beam direction was used to measure Rutherford scattering events for normalization. The scattered $\alpha$ particles were identified from the $\Delta$E vs E correlation plot. Figure~1 shows a typical two-dimensional $\Delta$E - E scatter plot from the detector telescope at backward angle
for $E_{lab}$ = 22 MeV. The Z = 2 events correspond to the elastic scattering of $^{4}$He and the unresolved $^{232}$Th inelastic excitations. In the analysis, quasi-elastic-scattering was defined as the sum of elastic plus inelastic events. An excitation energy window of 4.0 MeV in the scattered alpha energy spectrum is taken as the quasi-elastic events as shown by the rectangular box in Fig.~1. The energy window of 4.0 MeV was taken to include most of the low lying states of the $^{232}$Th nucleus. In Fig.~1 a lot of protons and low-energy $\alpha$ particles are also observed. These events may come from the reaction with the light element impurities, such as $^{12}$C and $^{16}$O, that may be present in the target.
The possible contribution of the evaporation $ \alpha $'s coming from the  $^{4}$He + $^{232}$Th compound system is found to be negligible from the PACE2~\cite{14} calculations for the present incident energy range. The ratio of quasi-elastic cross section to the Rutherford cross section was obtained by dividing the corresponding number of counts in the alpha particle band of the $\Delta$E - E spectrum by the number of elastic events in the monitor. The ratios were normalized to unity at the energies well below the coulomb barrier. The normalized ratio gives the differential quasi-elastic cross section relative to the Rutherford scattering cross section.
The quasi-elastic excitation function at the angle 160$^\circ$ as shown in Fig.~2 is used to determine the representation of fusion-barrier distribution $D_{qel}$(E, 160$^\circ$) using Eq.~(2). The quasi-elastic barrier distribution corresponding to  $D_{qel}$(E, 180$^\circ$) is obtained  from $D_{qel}$(E, 160$^\circ$) by appropriate centrifugal energy correction~\cite{7}.\\

 \begin{figure}
\includegraphics[scale=0.59]{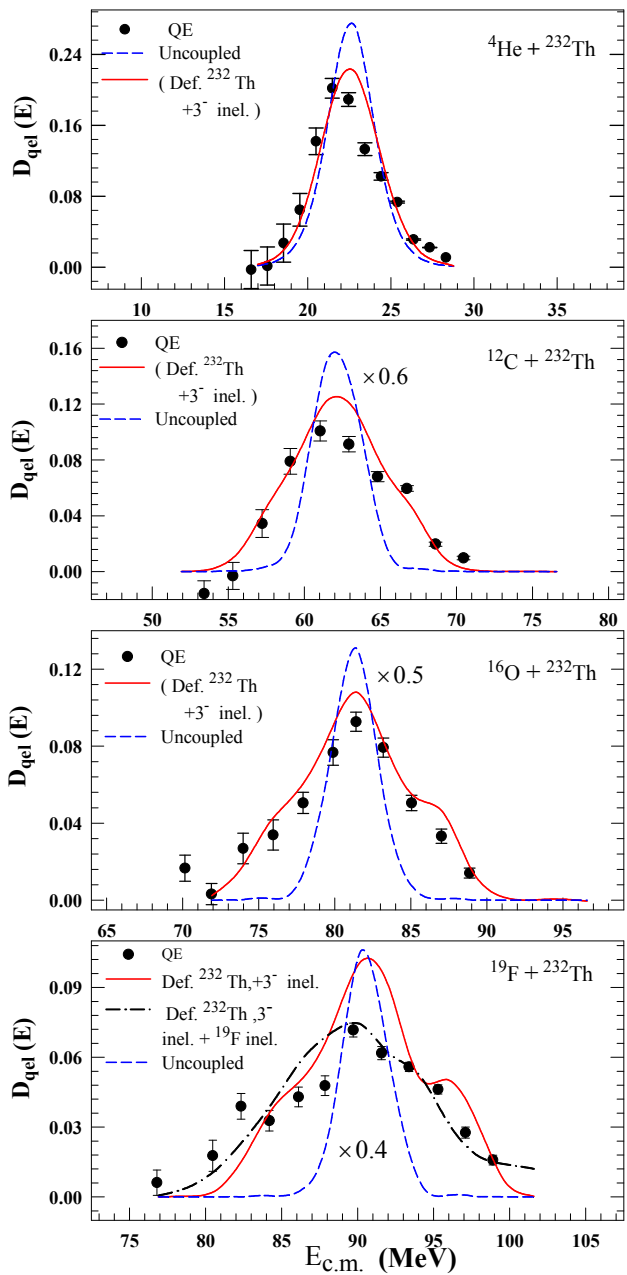}% Here is how to import jpg/ png/ pdf figure
\caption{\label{fig3} (Color online) Fusion-Barrier distribution for $^{4}$He + $^{232}$Th, $^{12}$C + $^{232}$Th, $^{16}$O + $^{232}$Th and $^{19}$F + $^{232}$Th systems.
The dashed lines represent the uncoupled barrier distributions. The continuous, and dashed-dot lines are the result of CCDEF calculations including channel couplings due to only target states and both target and projectile states respectively. }
\end{figure}

%\section{\label{sec:level3} Results and Discussion}
The experimental representation of barrier distribution for $^{4}$He + $^{232}$Th system along with  $^{12}$C + $^{232}$Th, $^{16}$O + $^{232}$Th and $^{19}$F + $^{232}$Th systems is shown in Fig.~3. The continuous and dashed lines in the figure are results of the coupled channel fusion model calculations using the code CCDEF~\cite{15}. The fusion excitation functions obtained from CCDEF calculations were converted into the fusion barrier distributions using Eq.~(4). The normalization values of (${\pi}{{R_{f}}^2}$) are determined from the relation,
\begin{equation}
T_{0}= \left(\frac{1}{\pi {R_f}^2}\right) \frac{d}{dE}[E \sigma_{fus}(E)],
\end{equation}
where $T_0$ $\rightarrow$1 at energies well above the Coulomb barrier for various systems.\\

The CCDEF calculations were performed for the $^{4}$He + $^{232}$Th system, including  couplings of the ground-state deformation of the $^{232}$Th target with
deformation parameters $\beta_2$, $\beta_4$, and the inelastic excitation of the 3$^{-}$ state at energy 0.774 MeV. 
The value of $\beta_2$ was taken to be 0.26 from the literature~\cite{16}. The $\beta_4$ and $\beta_3$ deformation parameters were varied in the calculation to get the best fit to the experimental data.
The experimental result on fusion-barrier distribution for  $^{4}$He + $^{232}$Th is well reproduced with $\beta_2$ = 0.26 and $\beta_3$ = 0.17 without including $\beta_4$ deformation in the CCDEF calculations. It may be noted that other combinations of $\beta_4$ and $\beta_3$ can also be used to reproduce the experimental barrier distribution for the $^4$He + $^{232}$Th system, but with the same combinations it is not possible to explain simultaneously the barrier distributions for the other systems consistently.\\
 
In Fig.~3, the dashed lines represent the uncoupled barrier distributions. The continuous lines are the result of the CCDEF calculations considering the coupling of static deformation with $\beta_2$ = 0.26 and  3$^{-}$ inelastic state at energy 0.774 MeV with $\beta_3$=0.17 of the target, as mentioned earlier. It can be seen that for the $^{4}$He + $^{232}$Th reaction the measured barrier distribution is quite similar to that of the uncoupled one because of the smaller value of the $Z_pZ_t$ product, due to which the coupling strength is very small~\cite{17}. It is also observed that for the reactions involving $^{4}$He, $^{12}$C, and $^{16}$O projectiles the experimental data are well explained by the continuous curve by considering only the channel couplings due to target intrinsic states. However, for the $^{19}$F + $^{232}$Th system the continuous line does not match with the experimental data. This is because of the presence of various low lying excited states in case of  $^{19}$F nucleus, which influence the fusion process. In order to explain the experimental representation of fusion-barrier distribution  for the $^{19}$F + $^{232}$Th
system, along with the target channel couplings the following inelastic states of the  projectile at 0.197, 1.346, 1.544, and 2.780 MeV with the deformation parameters, $\beta_2$ = 0.55, $\beta_3$ = 0.33, $\beta_2$ = 0.58 and $\beta_4$ = 0.22~\cite{18}, respectively are required to be included in the CCDEF calculation. These values of the deformation parameters of $^{19}$F used to fit the experimental data agree quite well to those obtained from inelastic excitation of $^{19}$F by (d,d$^\prime$),(p,p$^\prime$), and ($\alpha $,$\alpha ^\prime$) reactions~\cite{19,20,21}. The dashed-dot curve in Fig.~3, corresponding to the channel couplings of both target and projectile excited states, fits the experimental data for $^{19}$F + $^{232}$Th reasonably well. In order to demonstrate the effect of coupling of various inelastic states of $^{19}$F on fusion-barrier distribution, coupled-channel calculations have been carried out by including various inelastic couplings of $^{19}$F one-by-one; it is observed that experimental data are well explained if we include four low lying inelastic states of $^{19}$F along with the target state couplings, as shown in Fig~4.
It may be noted that coupled-channel fusion model calculation by the code CCFULL~\cite{22} is considered to be more accurate as it takes into account couplings to all order, whereas in the case of CCDEF linear coupling, approximation is used. In the present work, we have used the CCDEF code for fusion-barrier distribution calculations due to the limitations of CCFULL to include couplings of more than two modes of excitations for  target/projectile  in the calculation. In order to investigate the sensitivity of channel couplings in these two codes, a comparative study of fusion-barrier distribution
predictions of CCDEF and CCFULL has been carried out by considering coupling of various combination of two inelastic excitations of $^{19}$F at a time along with target state couplings. It is observed that predictions of fusion-barrier distributions by CCFULL and CCDEF codes  are similar for couplings of  the first two low-lying inelastic states (${\frac{5}{2}}^{+}$, ${\frac{5}{2}}^{-}$) of $^{19}$F. But  for  the inclusion of couplings of higher than two inelastic states (${\frac{3}{2}}^{+}$, ${\frac{9}{2}}^{+}$) of $^{19}$F the predictions of CCFULL and CCDEF show some differences. Particularly, CCFULL predicts more prominent structures in barrier distribution in comparison to CCDEF for the $^{19}$F + $^{232}$Th system. The inclusion of  couplings of the first two low lying inelastic states of $^{19}$F in the CCDEF calculations grossly describe the experimental fusion barrier distribution of the $^{19}$F + $^{232}$Th system as shown in Fig~4; by including couplings of all four low-lying inelastic states, the comparison between experiment and calculation improves.   
\\ 
\begin{figure}
\includegraphics[scale=0.4]{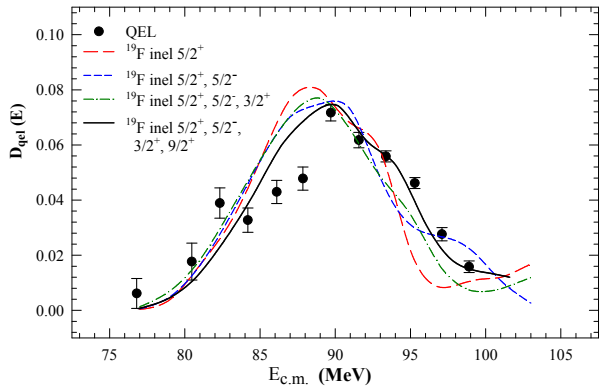}% Here is how to import jpg/ png/ pdf figure
\caption{\label{fig4} (Color online) Fusion-barrier distribution for the $^{19}$F + $^{232}$Th system. Various lines are the result of the CCDEF calculations including couplings of various inelastic states of $^{19}$F along with the coupling of target states (see text).}
\end{figure}

In order to investigate the dependence of barrier width as a function of projectile and target charge product ($Z_pZ_t$), the standard deviations ($\sigma _B$) of the measured barrier distributions were calculated from the relation, {$\sigma _B$ = ${\sqrt{<B^2> - {B_0}^2}}$}, where $B_0$ is the average barrier height. The width of the experimental barrier distribution (${\Delta}{B_{exp}}$) is  obtained from ${\Delta}{B_{exp}}$ = 2.35${\times}{\sigma _B}$, which is plotted as a function of the target projectile charge product $Z_pZ_t$ as shown in Fig.~5a. The correlation of ${\Delta}{B_{exp}}$/$B_0$, with $Z_pZ_t$ is shown in Fig.~5b. It can be seen that for $^{4}$He, $^{12}$C, and $^{16}$O + $^{232}$Th reactions, ${\Delta}{B_{exp}}$ increases linearly with $Z_pZ_t$ while ${\Delta}{B_{exp}}$/$B_0$ decreases systematically as a function of the same. For $^{19}$F + $^{232}$Th reaction both ${\Delta}{B_{exp}}$ and ${\Delta}{B_{exp}}$/$B_0$ deviate from the trends indicating the projectile structure effect for this reaction.\\\\

\begin{figure}
\includegraphics[scale=0.44]{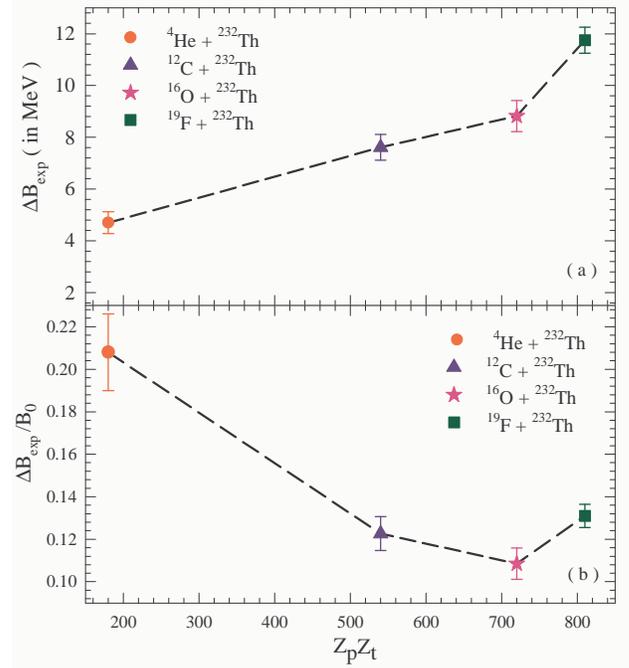}% Here is how to import jpg/ png/ pdf figure
\caption{\label{fig5} (Color online) (a) The experimental width of the fusion barrier distribution ${\Delta}{B_{exp}}$ and (b) ${\Delta}{B_{exp}}$/$B_0$ as a function of  the target projectile charge product. The dashed lines are guide to the eye.}
\end{figure}

%\section{Summary and Conclusions}

In summary, a systematic study of barrier distributions has been carried out for the $^{4}$He + $^{232}$Th, $^{12}$C + $^{232}$Th, $^{16}$O + $^{232}$Th and $^{19}$F + $^{232}$Th systems to investigate the relative importance of target and/or projectile couplings in explaining the measured barrier distributions. It is seen that the target deformation parameters $\beta_2$ and $\beta_3$ only are required to fit the experimental data for various projectiles. The role of the hexadecapole deformation parameter $\beta_4$ of $^{232}$Th is found to be less significant in explaining the measured barrier distributions. No projectile structure effect is observed on the fusion barrier distributions in $^{4}$He, $^{12}$C, and $^{16}$O + $^{232}$Th reactions. For the $^{19}$F + $^{232}$Th reaction,
  the experimental representation of fusion barrier distribution could only be explained by including inelastic couplings of the projectile in the CCDEF calculation. It is observed that the width of the barrier distribution increases with projectile and target charge product. This observation is consistent with the fact that, for a deformed nucleus the range of barrier heights is proportional to $Z_pZ_t\beta R$~\cite{4}, as well as the strength of the inelastic coupling also increases with $Z_pZ_t$~\cite{15}. The experimental barrier distribution width is observed to be higher for the $^{19}$F + $^{232}$Th system than expected from the $Z_pZ_t$ systematics observed for other reactions. This suggests that other than the target structure effects, the projectile structure is also playing a role in the fusion process in the case of the $^{19}$F + $^{232}$Th system.
 For the $^{4}$He + $^{232}$Th reaction, the measured barrier distribution is very close to that of the uncoupled barrier distribution due to lower $Z_pZ_t$ product, for which the coupling strength is less. The width of the barrier distribution normalized to the average barrier when plotted with respect to the $Z_pZ_t$ product shows interesting transition behavior related to the projectile structure effect.\\

\end{document}